\documentclass[preprint,showpacs,amsmath,amssymb]{revtex4}
\usepackage{latexsym}
\usepackage{amsmath}
\usepackage{amssymb}
\usepackage{amscd}
\usepackage{bm}
\usepackage{graphicx}
\usepackage{CJK}

\begin{document}
\begin{CJK}{GBK}{song}

\title{The ground state energy of the frustrated ferromagnetic spin chain near the transition point}
\author{Ren-Gui Zhu}
\email{rgzhu@mail.ahnu.edu.cn}\affiliation{College of physics and
electronic information, Anhui Normal University, Wuhu, 241000, P. R.
China}

\begin{abstract}
The one-dimensional quantum spin-1/2 model with nearest-neighbor
ferromagnetic and next-nearest-neighbor antiferromagnetic
interaction is considered. The Hamiltonian is first bosonized by
using the linear spin wave approximation, and then is treated by
using the Green's function approach. An integral expression of the
quantum correction to the classical ground state energy is derived.
The critical behavior of the ground state energy in the vicinity of
the transition point from the ferromagnetic to the singlet ground
state is analyzed by numerical calculation, and the result is
$-8\gamma^2$.
\end{abstract}


\maketitle
%**********************************************************************************
\section{Introduction}
Low-dimensional frustrated spin models have been intensively
investigated both theoretically and experimentally\cite{Diep,Mike}.
The one-dimensional (1D) quantum spin-1/2 model with
nearest-neighbor ferromagnetic interaction $J_1$ and
next-nearest-neighbor antiferromagnetic interactions $J_2$ has
attracted much attention in recent
years\cite{Lu,Hart,Dmitr1,Zink1,Rich,Dmitr2,Furu,Dmitr3,Zink2,Ende1}.
The growing interest was triggered by experimental studies of
various quasi-1D edge-sharing
cuprates\cite{Mizu,Masu1,Gibs,Gipp,Hase,Masu2,Ende2,Drech1,Capo,Drech2,Drech3,Drech4,Park,Male,Taru},
which can be described by this so-called 1D F-AF model. Its
Hamiltonian has a form:
\begin{equation}\label{eq01}
H=J_1\sum_{n=1}^N\left(\bm S_n\cdot\bm
S_{n+1}-\frac{1}{4}\right)+J_2\sum_{n=1}^N\left(\bm S_n\cdot\bm
S_{n+2}-\frac{1}{4}\right),
\end{equation}
where $J_1<0,J_2>0$, and the constant shifts $1/4$ is added to
secure the energy of the fully polarized state to be zero.

It is known that there is a ground state phase transition at the
point $\alpha\equiv J_2/|J_1|=1/4$\cite{Bade,Hama,Tone}. For
$\alpha<1/4$, the ground state is ferromagnetic. For $\alpha>1/4$,
the ground state is an incommensurate singlet with spiral spin
correlations. At $\alpha=1/4$, the ferromagnetic state is degenerate
with the singlet state.

One of the interesting problems in ground state properties is the
critical behavior of the ground state energy at this transition
point, which can be described by $E/N=-a \gamma^{\beta}$ with
$\gamma=\alpha-1/4$ and $0<\gamma<<1$. The coupled cluster method
gave $E/N\thicksim\gamma^2$\cite{Burs}. The perturbation theory
based on the classical approximation gave
$E/N=-4\gamma^2$\cite{Kriv}. However, by using Jordan-Wigner mean
field theory, Dmitriev and Krivnov\cite{Dmitr4} found recently that
the critical exponent $\beta$ should be less than $12/7$. Later,
they confirmed that $\beta=5/3$ by using scaling estimates of
perturbation theory\cite{Dmitr5}.

According to the previous results, it seems that the critical
behavior of the ground state energy in the vicinity of the
transition point is still a controversial problem, and more
elaborate methods are needed to obtain a definitive result. As an
important method in quantum magnetism, the spin wave theory often
gives out significant reference results. To our knowledge, however,
the spin wave theory in a regular way has not been applied to this
problem. In this paper, we try to make up this blank. Compared with
previous work, our following treatment is easier and
straightforward, and the final results have also some difference.

\section{The bosonization of the Hamiltonian}\label{sec2}
In the classical picture of the ground state of the 1D F-AF model,
the spins are vectors which form the spiral structure with a pitch
angle $\varphi$ between neighboring spins in the $xy$ plane, and all
spin vectors have the same canted angle $\theta$ from the $z$ axis.
We choose a reference state for the spin wave theory with all spin
vectors point along the $z$ axis. So we define The new spin
operators $\bm \eta_n$'s, which are related to the original ones by
the following rotational transformation:
\begin{equation}
\bm S_n=R_z(n\varphi)R_y(\theta)\bm\eta_n,
\end{equation}
where $R_y(\theta)$ is the rotational operator about the $y$ axis by
an angle $\theta$, and $R_z(n\varphi)$ is the rotational operator
about the $z$ axis by an angle $n\varphi$.

Using the new spin operators, the Hamiltonian (\ref{eq01}) can be
rewritten as
\begin{eqnarray}
H&=&\sum_{m=1}^2\sum_{n=1}^NJ_m\left[F_{++}(\theta,m\varphi)(\eta_n^+\eta_{n+m}^++\eta_n^-\eta_{n+m}^-)\right.\nonumber\\
&&+F_{+-}(\theta,m\varphi)\eta_n^+\eta_{n+m}^-+F_{+-}^*(\theta,m\varphi)\eta_n^-\eta_{n+m}^+
+F_{zz}(\theta,m\varphi)\eta_n^z\eta_{n+m}^z\nonumber\\
&&+F_{+z}(\theta,m\varphi)(\eta_n^+\eta_{n+m}^z+\eta_n^z\eta_{n+m}^-)\nonumber\\
&&\left.+F_{+z}^*(\theta,m\varphi)(\eta_n^-\eta_{n+m}^z+\eta_n^z\eta_{n+m}^+)\right]-\frac{N(J_1+J_2)}{4},
\end{eqnarray}
where $\eta_n^{\pm}=\eta_n^x\pm i\eta_n^y$, and the coefficient
functions are
\begin{gather}
F_{++}(\theta,\varphi):=\frac{1}{4}\sin^2\theta(1-\cos\varphi),\\
F_{+-}(\theta,\varphi):=\frac{1}{4}(\cos\varphi\cos^2\theta+\sin^2\theta+\cos\varphi-2i\sin\varphi\cos\theta),\\
F_{zz}(\theta,\varphi):=\cos\varphi\sin^2\theta+\cos^2\theta,\\
F_{+z}(\theta,\varphi):=[\frac{1}{4}\sin2\theta(\cos\varphi-1)-\frac{i}{2}\sin\varphi\sin\theta].
\end{gather}

Taking the linear spin-wave  approximation\cite{Hols}:
$\eta_n^+=\sqrt{2S}a_n, \eta_n^-=\sqrt{2S}a_n^{\dagger},
\eta_n^z=S-a_n^{\dagger}a_n$, and then the Fourier transformation:
$a_n=\frac{1}{\sqrt{N}}\sum_kb_ke^{ikn}$, we obtain the bosonic
Hamiltonian:
\begin{equation}\label{eq04}
H=E_0(\theta,\varphi)+\sum_kA(\theta,\varphi,k)b_k^{\dagger}b_k+\sum_kB(\theta,\varphi,k)(b_kb_{-k}+b_k^{\dagger}b_{-k}^{\dagger}),
\end{equation}
where
\begin{gather}
A(\theta,\varphi,k)=\sum_{m=1}^24SJ_m\mbox{Re}[F_{+-}(\theta,m\varphi)\cos mk]-\sum_m2SJ_mF_{zz}(\theta,m\varphi),\\
B(\theta,\varphi,k)=\sum_{m=1}^22SJ_mF_{++}(\theta,m\varphi)\cos mk,\\
E_0(\theta,\varphi)=N[\sum_{m=1}^2S^2J_mF_{zz}(\theta,m\varphi)-\frac{1}{4}(J_1+J_2)].
\end{gather}

We note here that the terms containing four Bose-operators have been
neglected as usually. The terms containing odd number of Bose
operators have been neglected as well according to the treatment in
Ref.\cite{Dmitr4}, because they have no contribution to the energy
in the mean-field approximation. It is only based on these
approximations that we can obtain the above regular bosonic
Hamiltonian, and its effectiveness can be demonstrated by the
results derived from it.

\section{The critical behavior of the ground state energy}\label{sec3}
In the following ,we take $S=1/2,J_1=-1$ and $J_2=\alpha$. The
energy function $E_0(\varphi,\theta)$ is minimized at
\begin{equation}
\varphi=\cos^{-1}\frac{1}{4\alpha}, \ \theta=\pi/2
\end{equation}
The minimum of $E_0(\varphi,\theta)$ is just the classical ground
state energy:
\begin{equation}
\epsilon_{cl}\equiv\frac{E_{cl}}{N}=-\frac{(\alpha-1/4)^2}{2\alpha},
\end{equation}
which gives the critical behavior: $\epsilon_{cl}=-2\gamma^2$ as
$\gamma\to0$.

In order to obtain the quantum correction to the classical ground
state energy, we use the double-time Green's function
approach\cite{Tyab,Frob} to derive the zero-temperature average
$\langle b_k^{\dagger}b_k\rangle_0$, $\langle b_kb_{-k}\rangle_0$
and $\langle b_k^{\dagger}b_{-k}^{\dagger}\rangle_0$. By solving the
group of equations of motion for the four Green's functions:
$\langle\langle b_k;b_k^{\dagger}\rangle\rangle_{\omega}$,
$\langle\langle
b_{-k}^{\dagger};b_k^{\dagger}\rangle\rangle_{\omega}$,
$\langle\langle b_{-k};b_k\rangle\rangle_{\omega}$ and
$\langle\langle b_k^{\dagger};b_k\rangle\rangle_{\omega}$, and using
the spectral theorem and taking the zero-temperature limit, we
finally get:
\begin{equation}\label{eq02}
\langle
b_k^{\dagger}b_k\rangle_0=\frac{1}{2}\left(\frac{A}{C}-1\right),
\end{equation}
and
\begin{equation}\label{eq03}
\langle b_kb_{-k}\rangle_0=\langle
b_k^{\dagger}b_{-k}^{\dagger}\rangle_0=-\frac{B}{C},
\end{equation}
where $C=\sqrt{A^2-4B^2}$ is the pole of the above green's
functions.

By taking Eq.(\ref{eq02}) and Eq.(\ref{eq03}) into the bosonic
Hamiltonian (\ref{eq04}), we obtain the energy function per site
including the quantum correction:
\begin{equation}\label{eq05}
\epsilon=\epsilon_0-\frac{1}{2\pi}\int_{k\in
D}dk(A-\sqrt{A^2-4B^2}),
\end{equation}
where $\epsilon=\frac{E}{N}$, $\epsilon_0=\frac{E_0}{N}$, and the
quasimomentums $k$ is restricted in the region $D=\{k|0\le
k\le\pi\mbox{ and }A^2-4B^2\ge0\}$.

The integral expression on the right side of Eq.(\ref{eq05}) is the
quantum correction to the classical energy. Based on this expression
the critical behavior of the total ground state energy can be
analyzed as follows. Considering the values of $\theta$ and
$\varphi$ in the classical ground state, we take $\theta=\pi/2$ in
Eq.(\ref{eq05}) and obtain the energy function
$\epsilon(\varphi,\alpha)$. Furthermore, from the relation
$\cos\varphi=\frac{1}{4\alpha}$, we get $\varphi=\sqrt{8\gamma}$ for
$\gamma\to0$. Finally, the critical behavior of the total ground
state energy can be described by the function:
\begin{equation}\label{eq06}
\frac{E(\gamma)}{N}=\epsilon(\sqrt{8\gamma},\gamma+\frac{1}{4}),
\mbox{ for }0<\gamma<<1.
\end{equation}

\begin{figure}
   \begin{center}
   \setlength{\unitlength}{1cm}
     \begin{picture}(12,8)
     \put(0,0){\includegraphics[scale=1.0]{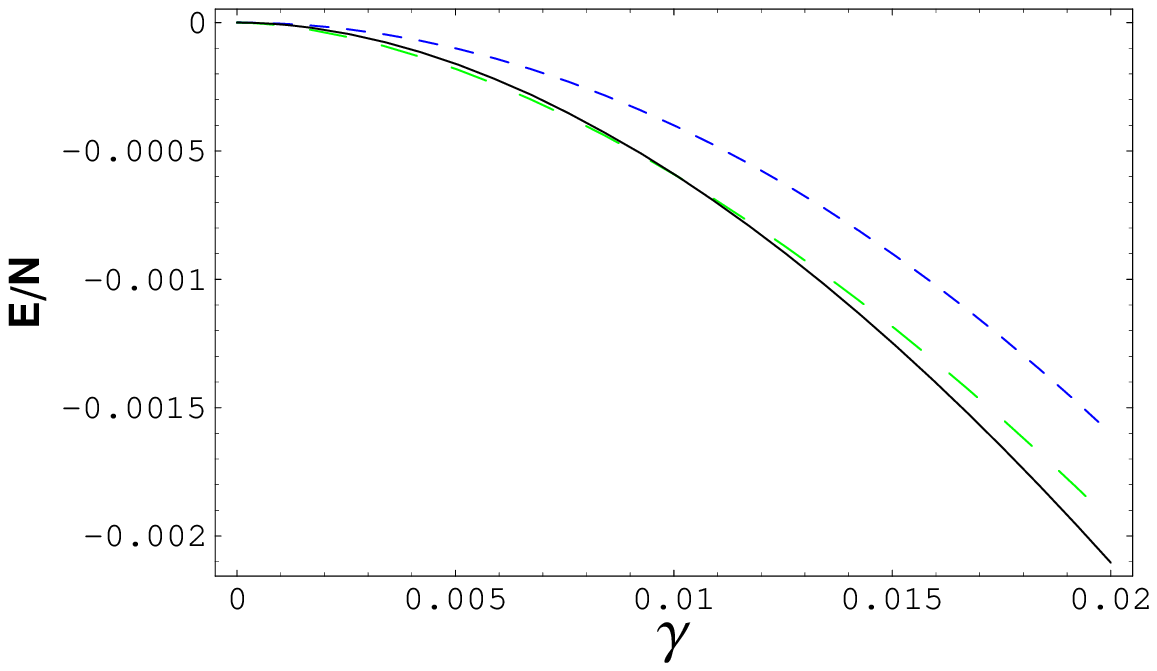}}
     \put(1.5,2.5){\setlength{\unitlength}{0.5cm}
                \begin{picture}(10,8)
                       \put(1.5,-2){\includegraphics[scale=0.5]{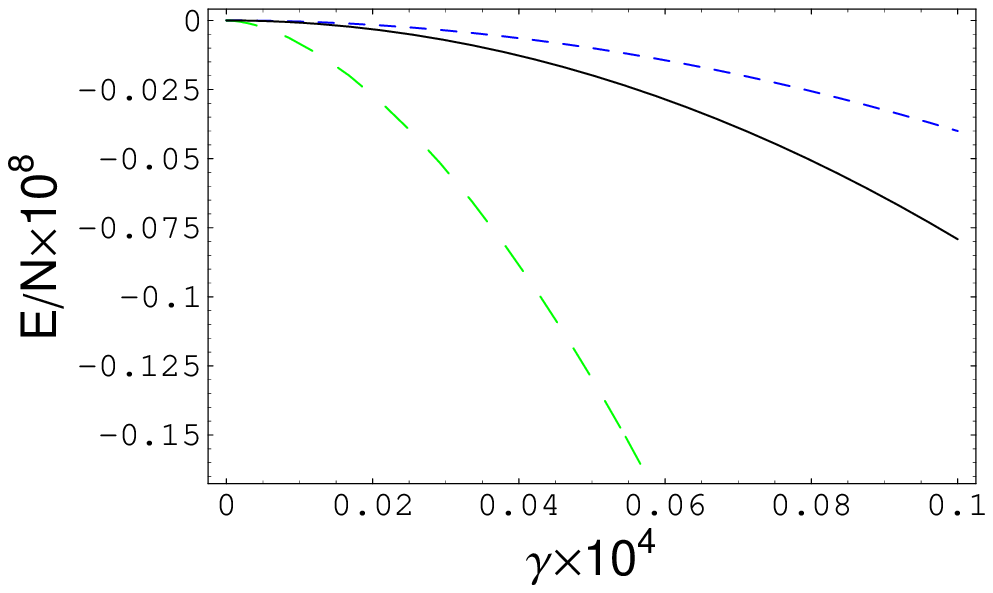}}
                \end{picture}
                }
     \end{picture}
     \caption{\label{fig1}(Color online) The dependence of the ground state energy on
     $\gamma$ given by Eq.(\ref{eq06}) (solid line), $E/N=-4\gamma^2$ of Ref.\cite{Kriv} (short-dashed line),
     and $E/N=-1.585\gamma^{12/7}$ of Ref.\cite{Dmitr4} (long-dashed line).}
   \end{center}
\end{figure}

\begin{figure}
  \begin{center}
  \includegraphics[scale=0.8]{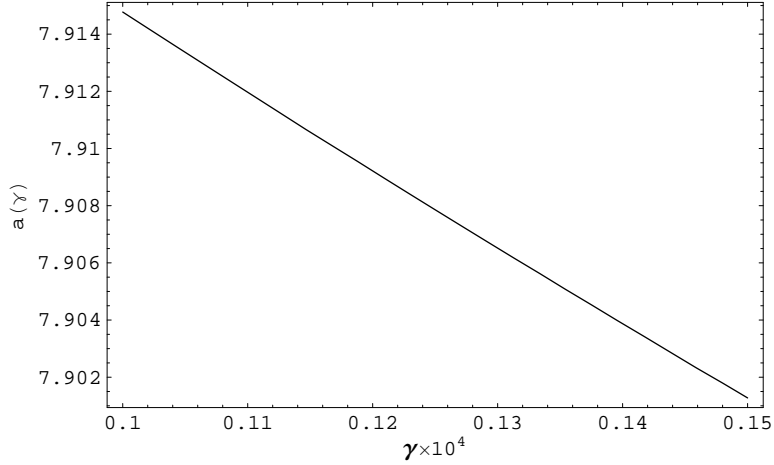}
  \end{center}
  \caption{\label{fig2}The curve of the function $a(\gamma)$ calculated from $E(\gamma)/N=-a\gamma^2$ predicts the limit value
  $a(\gamma\to0)\approx8.0$}.
\end{figure}

Fig.\ref{fig1} shows that in the quite low region of $\gamma$ our
result calculated from Eq.(\ref{eq06}) is close to $-4\gamma^2$ of
Ref.\cite{Kriv}, While in the most part of the region
$\gamma\in(0,0.02)$ our result is close to $-1.585\gamma^{12/7}$ of
Ref\cite{Dmitr4}. So from the calculation result, we can predict
that the critical behavior of the ground state energy given by
Eq.(\ref{eq06}) is $E(\gamma)/N=-a\gamma^{2}$. The coefficient $a$
is given by $a(\gamma)=-E(\gamma)\gamma^{-2}/N$ with the limit
$\gamma\to0$. From Fig.\ref{fig2}, one can predict that
$a\approx8.0$.

On the other hand, doing a regular expansion in powers of small
parameter $\varphi$ to the fourth order, and taking $\alpha\to1/4$,
we obtain $A-C=F_2(k)\varphi^2+F_4(k)\varphi^4$ with the
coefficients:
\begin{equation}
F_2(k)=0,\ F_4(k)=\frac{(\cos2k-\cos k)^2}{8(3-4\cos k+\cos 2k)}.
\end{equation}
The quantum correction to the classical part of the ground state
energy is
\begin{equation}
-\frac{\varphi^4}{2\pi}\int_0^{\pi}F_4(k)dk=-\frac{3\varphi^4}{32}=-6\gamma^2,
\mbox{ for }\varphi=\sqrt{8\gamma},\gamma\to0,
\end{equation}
which is consistent with the above numerical calculation. However,
if we do the expansion to the sixth order, the integral for the
coefficient $F_6$ is infrared-divergent. This fact may implicit the
restriction on the effectiveness of our treatment to this model.
Nevertheless, our calculations and results derived from
Eq.(\ref{eq05}) and Eq.(\ref{eq06}) seem reasonable, and the whole
treatment is straightforward and easier than previous treatments.
\section{Conlusion}
In this paper, The ground state energy of 1D F-AF model (\ref{eq01})
in the vicinity of the transition point $\alpha=1/4$ is considered.
Using the linear spin-wave approximation and the green's function
approach in a very regular way, we obtain an integral expression
Eq.(\ref{eq05}) for the ground state energy including the quantum
correction. The critical behavior of the ground state energy is
described by the function $\epsilon(\sqrt{8\gamma},\gamma+1/4)$ in
Eq.(\ref{eq06}), and the result is $E(\gamma)/N=-8\gamma^2$ which is
compared with the previous results in Fig.\ref{fig1}.

It is known that the total energy of the $M$-magnon state of the
model (\ref{eq01}) is $E_M=-8M\gamma^2$\cite{Kriv,Ovch}. The
critical behavior of the ground state energy $E/N=-8\gamma^2$ means
that $N$ noninteracting magnons are created in the ground state.

\section*{Acknowledgments}
This work belongs to the collegial key project of natural science
research funded by Anhui province in China under Grant No.
KJ2010A131, and is partly supported by the National Natural Science
Foundation of China under Grant No.10947138.

%**************************************************************************

\end{CJK}
\end{document}